\documentclass[12pt]{article}
\usepackage[scale=0.8]{geometry}

\usepackage{authblk}

\usepackage{lmodern}
\usepackage{epsfig}
\usepackage{float}
\usepackage{amscd}
\usepackage{amsmath, amsthm}
\usepackage{amssymb}
\usepackage{mathtools,commath}

\usepackage[font=small,labelfont=bf]{caption}
\usepackage{subcaption}
\usepackage{booktabs}

\usepackage{natbib}
\newcommand{\Reyuls}{\mathbb{R}}
\newcommand{\bo}{{\partial\Omega}}
\newcommand{\bog}{\Gamma}

\newcommand{\sdiv}{{{\nabla\cdot} \,}}
\newcommand{\du}{{\mathcal D}(u)}
\newcommand{\dv}{{\mathcal D}(v)}

\begin{document}

\title{
Assessment of Numerical Lift Coefficient Data for a Circular Cylinder with Application to Bladeless Turbines
}
\date{Draft: \today}

\author{L. Ridgway Scott}
\affil{Department of Mathematics, University of Chicago,\\Chicago, Illinois 60637, USA}
\author{Henry von Wahl}
\affil{Institut für Mathematik, Friedrich-Schiller-Universität Jena,\\ Inselplatz 5, 07743 Jena, Germany}

\maketitle

\begin{abstract}
We assess computed lift coefficient data for flow past a circular cylinder to evaluate their suitability for practical applications. Specifically, we consider lift coefficient data for a circular cylinder over Reynolds numbers from 120 to 8000. The results are obtained from two-dimensional finite element simulations of the incompressible Navier–Stokes equations using pressure robust discretizations. We compare the computed lift coefficients with published experimental and numerical results, finding good agreement in some cases but significant disagreement in others. Because lift fluctuations are central to vortex-induced vibration concepts, these data therefore provide input for the analysis and preliminary design of bladeless turbines.
\end{abstract}

\section{Introduction}

There has been considerable interest in the drag coefficient for the flow around a cylinder;
see \cite{lrsBIBkl} for recent results and extensive references to previous work.
However, there has been less attention to lift data \cite{ref:FloresCelisBlancosubsoncyl},
although in many ways the oscillatory lift has a significant possibility to contribute
to industrial applications such as bladeless energy generation \cite{lrsBIBkt}.
Even for fixed cylinders, the fluctuations in lift are an order of magnitude greater than
the fluctuations in drag. For the case of rotating cylinders, we refer to
\cite{lrsBIBei,ref:Magnusreviewseifert}, where the flow around rotating cylinders has
been studied with regard to the Magnus effect.

In the previous paper \cite{lrsBIBkl}, we 
focused on the more commonly inspected aspect of the resulting drag coefficient.
However, lift is the key quantity of interest when considering wind energy generation.
Thus our attention turned 90 degrees (pun intended) and, we realized that the
data on lift was more sparse and contradictory.
The main focus of the current paper is to  put  more emphasis on lift,
both computationally and experimentally.
We are in the process of designing experiments to measure lift.

We review computations done in \cite{lrsBIBkl} for which the drag
coefficient was the main concern, focusing instead on lift computations. In \cite{lrsBIBkl}, the main focus was on the transition from a periodic flow to a chaotic one as the Reynolds number increases, based on a number of advanced finite element discretizations of the incompressible Navier-Stokes equations. 
Here, we will see that there is reasonable agreement between the lift data from the best and most reliable performing methods in \cite{lrsBIBkl} and other computational and experimental studies of lift.
But we will also see some discrepancies, as was found with drag data in \cite{lrsBIBkl},
and this suggests that vibrating cylinders should be considered as a contributor.

The remainder of this work is structured as follows.
In Section~\ref{sec.equations}, we cover the equations under consideration.
Section \ref{sec.setup} covers the concrete set up used for the numerical simulations
reviewed here, and briefly outlines the finite element methods used.
In Section~\ref{sec.discussion}, we review the computed lift data with respect
to results available in the literature, and in Section~\ref{sec:industimpl},
we discuss how these results may be applied in the design of bladeless windturbines.
Section \ref{sec:prestud} reviews what is known about computations allowing
cylinder vibrations and research on bladeless generators.
We then conclude with some final remarks in Section~\ref{sec:conclusions}.

\section{Setting the problem and model equations}
\label{sec.equations}

Let $(u,p)$ be the solution of the time-dependent Navier-Stokes
equations in a domain $\Omega\subset\Reyuls^d$ containing an obstacle with
boundary $\Gamma\subset\bo$. This fulfills the equations
\begin{subequations} \label{eqn:firstnavst}
\begin{align}
    && \partial_t u-\nu\Delta  u +  u\cdot\nabla u + \nabla p &= 0 &&\text{in}\;\Omega,\\
    && \sdiv u &=0&&\text{in}\;\Omega,
\end{align}
\end{subequations}
with the kinematic viscosity $\nu$, and together with boundary conditions
\begin{subequations}\label{eqn:bceesnavst}
\begin{align}
    u=g&\text{ on }\partial\Omega\backslash\Gamma\\
    u=0&\text{ on }\Gamma.
\end{align}
\end{subequations}
For the question of well-posedness of these equations, we refer to \cite{giraultraviart}.
We note that it can be confusing that the vortex street is not unique. 
There are two solutions, depending on which side of the cylinder the vortex starts.
But this is a simple example of a Hopf bifurcation \cite{Hopf1950berDA,ref:cylinderHopfvortexstreet}.
However this nonuniqueness is not in time evolution, but rather in the varying of the 
Reynolds number.
For example, there is a steady flow around a cylinder at any Reynolds number,
 however, this solution becomes unstable around Reynolds number 50.
For larger Reynolds numbers, if we start with a small perturbation on one side
of the cylinder, the vortex street develops predictably on one side of the cylinder.
If we switch the perturbation to the other side, 
the vortex street develops predictably on the other side of the cylinder.

\subsection{Weak formulation of the Navier-Stokes equations}
We will consider a finite element discretizations of \eqref{eqn:firstnavst} -- \eqref{eqn:bceesnavst}.
To this end, we observe that the Navier-Stokes equations can be written in a weak
(or variational) form as follows: Find $(u,p) \in H^1_g(\Omega)\times L^2_0(\Omega)$,
such that
\begin{equation*}
    (\partial_t u,v)_{L^2(\Omega)} + a(u, v) +c(u,u,v) + b(v, p) + b(u, q) = 0
\end{equation*}
for all $(v,q)\in (H^1_0(\Omega))^d\times L^2_0(\Omega)$.
The space $H^1_g(\Omega)$ is the space of vector-valued $H^1$ functions with
trace $g$ on the boundary, and $H^1_0(\Omega)$ is the space of $H^1$
functions with trace zero on the boundary. The space $L^2_0(\Omega)$ is the space of
$L^2$ functions with mean zero. The bilinear forms $a(\cdot, \cdot)$ and
$b(\cdot, \cdot)$ are defined by
\begin{align*}
    a(u, v) \coloneqq \int_\Omega \frac{\nu}{2} \mathcal{D}(u):\mathcal{D}(w)\dif x
    \qquad\text{and}\qquad
    b(v, q) \coloneqq -\int_\Omega q (\nabla \cdot v) \dif x,
\end{align*}
respectively, where $\dv=\nabla v+\nabla v^t$ and the colon indicates the
Frobenius inner-product of matrices. The convective (non-linear) term takes the form
\begin{equation*}
    c(u,v,w) \coloneqq \int_\Omega (u\cdot\nabla v)\cdot w  \dif x .
\end{equation*}

\subsection{Drag and lift evaluation}

The drag and lift are given by the formula
\begin{equation*}
    \beta(v) = \int_{\bog} \big((\nu\du - pI)n\big)\cdot v \dif s,
\end{equation*}
where $I$ is the identity matrix.
The drag is obtained using $v=(1,0)$ and the lift using $v=(0,1)$.
Note that while the evaluation of this boundary integral is possible, it is also possible to test the residual
with a non-conforming test function, see \cite[Remark D.2]{JohnBook16} and \cite{ref:refvalcyliftdragVolkerJohn}.
This is more accurate and stable than the direct evaluation of the surface integral.
Indeed, the volumetric evaluation has double the order of accuracy, see \cite{braack2006solutions}
for the proof in the steady case. 
We will use both approaches depending on the finite element method used,
see Section~\ref{sec.computational-methods} below.

\begin{figure}
    \centering
    \begin{subfigure}{0.4\textwidth}
        \includegraphics[height=5cm]{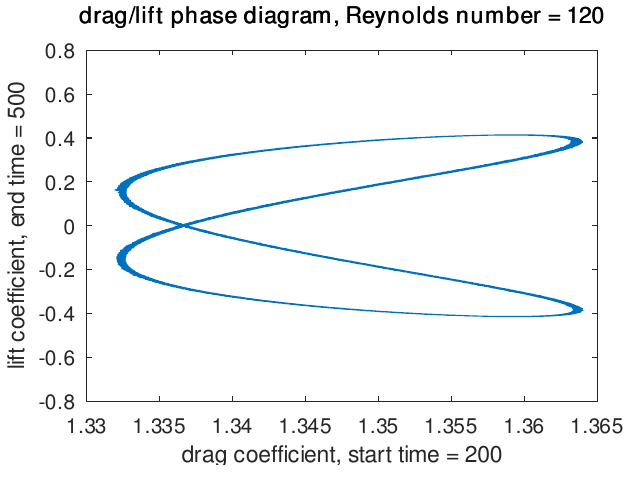}
        \caption{Re=120}
    \end{subfigure}
        \begin{subfigure}{0.4\textwidth}
        \includegraphics[height=5cm]{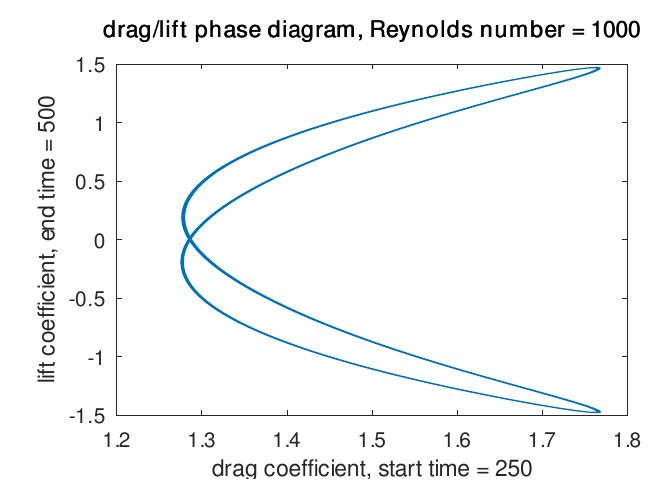}
        \caption{Re=1000}
    \end{subfigure}
    \caption{Lift-drag phase diagram for different Reynolds numbers resulting in a periodic flow.
Computed using the MCS finite element method.}
    \label{fig:fazedataRe}
\end{figure}

\section{Problem Set-up}
\label{sec.setup}
The choice of computational domain follows \cite{lrsBIBki,lrsBIBkl}.

\subsection{Geometry} 
We consider the domain
\begin{equation*}
    \Omega=\{(x,y):-30<x<300,\,|y|<30,\;x^2+y^2>1\}.
\end{equation*}
The boundary condition is set as $g = (1, 0)^T$ on the outer boundary and $g=(0,0)$ on the cylinder.
The cylinder diameter is the reference length, so the Reynolds number is given by $Re = 2 / \nu$.
We consider the time interval $[0, 500]$.

\begin{figure}
    \centering
    \begin{subfigure}{0.4\textwidth}
        \includegraphics[width=2.4in,angle=0]{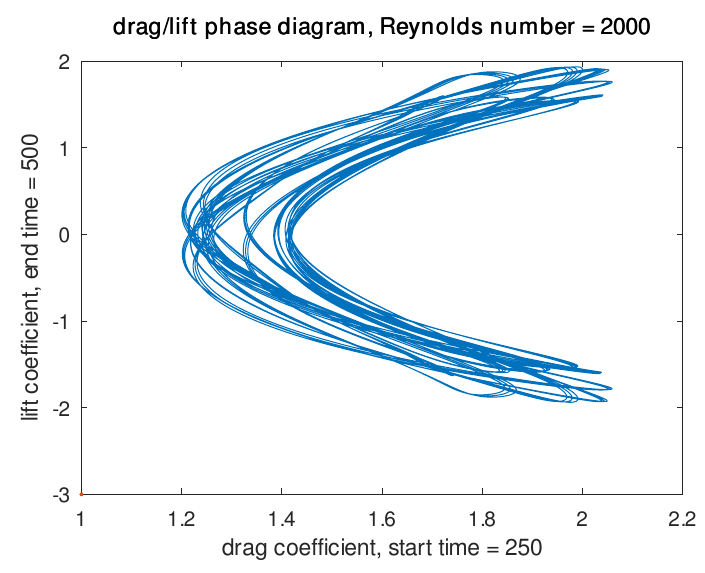}
        \caption{MCS}
    \end{subfigure}
    \begin{subfigure}{0.4\textwidth}
        \includegraphics[width=2.4in,angle=0]{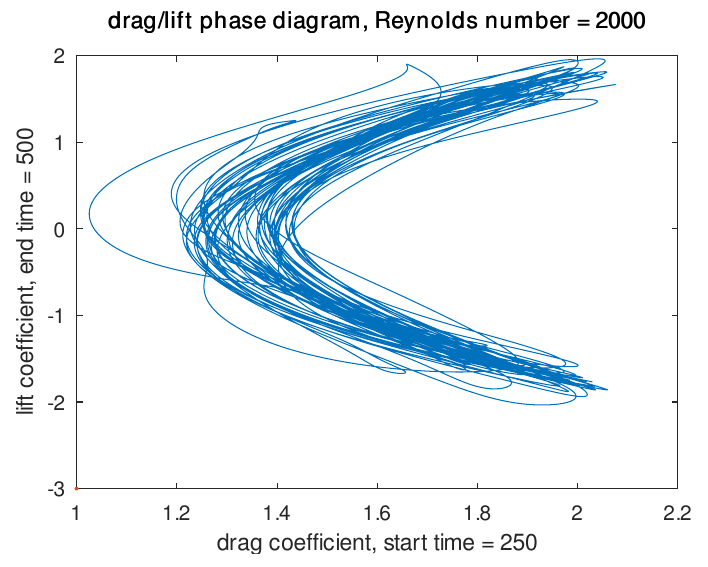}
        \caption{SV}
    \end{subfigure}
    \caption{Lift-drag phase diagram for Re=2000.}
    \label{fig:fazedatacomp}
\end{figure}

\subsection{Computational methods}
\label{sec.computational-methods}

There were many computational methods examined in \cite{lrsBIBkl}, but two stood out as providing the most accuracy.
One of them was the Scott--Vogelius method \cite{lrsBIBib} using conforming piecewise quartic velocities and
discontinuous cubic pressure approximations.
On the finest mesh considered, the resulting finite element space has 551434 degrees of freedom.
For the temporal discretization, the fully implicit second order Crank-Nicolson scheme was used
with a time-step $\Delta t=0.01$, which we found to sufficient to discretize in time.
Drag and lift are computed using both the boundary and volume formulation. The results presented
are for the more accurate volume formulation.
We will refer to this method as SV for short.

Another method was an $H(div)$-conforming method from \cite{gopalakrishnan2020weak}, using nonconforming piecewise quartic
velocities, together with the upwind formulation of the convective term. 
The latter method also was exactly mass conserving and based on a mixed stress formulation for the Stokes part.
To avoid the saddle point problem, a penalty formulation is used here. Consequently, we do not solve for the pressure (which would also be a discontinuous cubic) implicitly but reconstruct this from the velocity approximation. The stress and velocity space then have a total of 604047 degrees of freedom on the finest mesh. We then use the boundary formulation of the drag and lift functional here, as the residual form does not apply here. In time, we discretize with a second order implicit-explicit time stepping scheme. Here, the diffusion part is treated implicitly, while convection is handled explicitly. This results in a CFL condition on the time-step, so we needed $\Delta t = 0.00025$ on the finest mesh.
We will refer to this method as MCS for short. 

Both of these methods result in pointwise divergence-free velocities, i.e., they conserve mass exactly and are pressure robust. The latter means that the velocity error is independent of the pressure error. Both methods were implemented in \texttt{NGSolve} \cite{schoeberl_ngsolve}.

The upwind form of the convective term was found to be critical in the MCS method, as this resulted is just enough numerical dissipation on under resolved meshes to capture the periodic results obtained after a sufficient number of mesh refinements and at low to moderate Reynolds numbers. Indeed, since the SV method does not contain any numerical dissipation, we found that the MCS was able to give reliable, mesh converged, results at higher Reynolds numbers. The same results were only obtained with SV once all scales where fully resolved.

\section{Quantities of interest}
\label{sec.discussion}

For Reynolds numbers between 50 and 1000, the flow evolves into the von Karman vortex
street \cite{lrsBIBkl} with a periodic drag-lift profile.
This is depicted in Figure \ref{fig:fazedataRe} for two Reynolds numbers, 120 and 1000.
What is plotted are the pairs $(x_n,y_n)=(d(t_n),\ell(t_n))$, where $d$ is the drag and
$\ell$ is the lift, for all values of $t_n\in [250,500]$ for
which the simulation has produced velocity and pressure profiles.
See \cite{lrsBIBkl} for more details.
At both of these Reynolds numbers, the drag-lift profile is periodic, after an initial
start-up phase.
Plotted in Figure \ref{fig:fazedataRe} is the drag-lift phase diagram starting at $t=250$
up to the final time $t=500$ for the simulations.

\begin{figure}
\vspace{-60mm}
\centerline{\includegraphics[width=5.6in,angle=0]{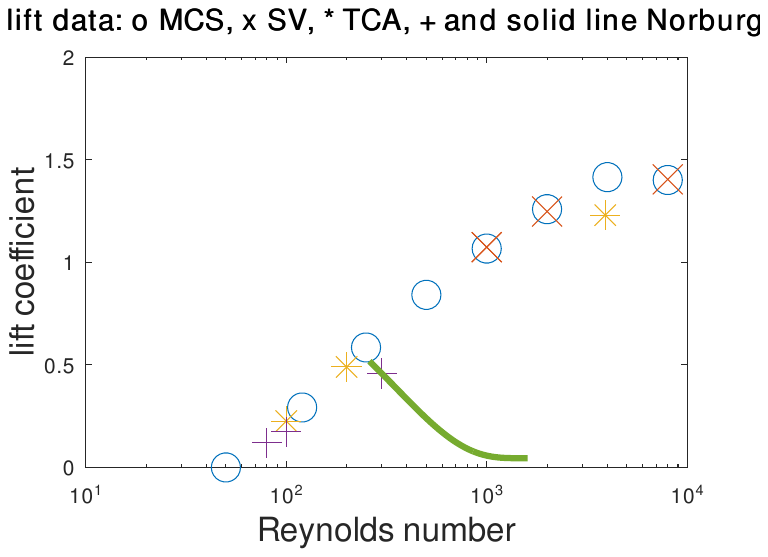}}
\vspace{-63mm}
    \caption{Lift data as a function of Reynolds number.
The o's were computed using the MCS scheme and the x's were computed using the SV scheme from \cite{lrsBIBkl}.
The asterisks correspond to lift data in the work of Tymchuk, Celis, and Alvarez 
(TCA) \cite[Table 5]{ref:FloresCelisBlancosubsoncyl}.
The data indicated by +'s and the solid line were taken from the work of
Norberg \cite{ref:norbergfluctlift}.}
    \label{fig:liftdata}
\end{figure}

By contrast, Figure \ref{fig:fazedatacomp} shows the chaotic nature of the drag-lift
phase diagram for Reynolds number 2000.
We see there the results of computations with the two different numerical methods
(SV and MCS) explained in detail in \cite{lrsBIBkl}.
Although both simulations are clearly chaotic, the two different numerical methods produce
slightly different phase diagrams, as would be expected with chaotic flow (small differences
quickly evolve differently since the Lyapunov exponent \cite{lrsBIBki} is positive).
On the other hand, the lift coefficient (the standard deviation of the lift data) are
quite similar, as indicated in Figure \ref{fig:liftdata}.

Figure \ref{fig:liftdata} shows significant agreement between the lift data corresponding to the computations
in \cite{lrsBIBkl} and the lift data in \cite[Table 5]{ref:FloresCelisBlancosubsoncyl}.
The latter considered numerical simulations using ANSYS with different turbulence models at high Reynolds numbers and some wind-tunnel experiments. Our results are in close alignment with their numerical results, which in turn are similar for all turbulence models considered. While in [18], no experimental lift data is recorded, there was good agreement between the wind tunnel experiments and numerical simulations for the drag coefficient, pressure coefficient and Strouhal data.

For Reynolds numbers up to a few thousand, the two-dimensional model is likely
to have significant agreement with three-dimensional data, at least for sufficiently
long cylinders compared to the cylinder diameter.
It is remarkable how good the agreement is between our computations and some of the
other data in Figure \ref{fig:liftdata}, even up to the point where it is known that
three-dimensional effects become important.
For flow past a cylinder, it is known \cite{ref:FazleHayakawatreedee} that the flow field
exhibits significant three-dimensional character at $Re=10000$.
We further caution that some three-dimensional effects may lead to spanwise
variation of lift along the cylinder \cite{ref:norbergfluctlift,ref:spanwisecylinder},
leading to reduced effective lift.

The data in Figure \ref{fig:liftdata} is substantially at variance
with \cite[Figure 2]{ref:norbergfluctlift}.
However, the latter contains a variety of computational and experimental data available
at the time, and this data shows a very large variance.
Even at low Reynolds numbers, a wide range of mean lift coefficients is given based
on two-dimensional simulations.
This is turn suggests that some simulations were not fully resolved.
Particularly around $Re=1000$, the reported root mean square (r.m.s.) lift coefficient
data has the largest variance, with values reported between 0.05 and 0.7.
We note that the drag-lift profile is still periodic at $Re=1000$, although several
simulations were not able to verify this due to a lack of accuracy and appropriate numerical dissipation
 as indicated in \cite{lrsBIBkl}.

Further data in \cite[Figure 3.36]{ref:liftexptsMIT} also disagrees with Figure \ref{fig:liftdata}.
There, the r.m.s.~sectional lift coefficient is reported between about 0.2 and 0.6 for
Reynolds numbers between $4\cdot10^4$ and $10^5$, with a trend of increasing lift
coefficient with increasing Reynolds number.
See also data in \cite[Figure 3.30]{ref:liftexptsMIT} where a comparison is made with other
experiments, including the work of Gerrard \cite{gerrard1965disturbance}
and others \cite{humphreys1960circular,kacker1974fluctuating,keefe1962investigation,
leehey1970aeolian,tanida1973stability}.
Here the variance is even greater with values reported between 0.03 and 0.6 for Reynolds number between $4\cdot10^4$ and $10^5$, and a less clear trend for the lift coefficient with increasing Reynolds numbers.
As in \cite{lrsBIBkl}, such discrepancies could be due to cylinder vibration in the experimental work.

A single data point of interest comes from \cite{ref:marineliftMIT}.
The subject of the thesis was vibrating cylinders, but there was a reference case
consisting of the drag and lift at Reynolds number $10^4$ for a fixed (non-vibrating) cylinder towed through water.
The distribution of drag and lift values over a series of experiments is given in \cite[Figures 3.2 and 3.3]{ref:marineliftMIT}. Furthermore, \cite[Figures 4.1 and 4.2]{ref:marineliftMIT} show that the data is not correlated with the event index, i.e., dismantling, rebuilding and recalibrating the experimental set-up did not have an effect on the reported numbers.
The mean over the conducted experiments of lift coefficient is 0.3842 with a standard deviation of 0.0873.
Nevertheless, the spread of reported lift values is wide and in particular wider than those
for the mean drag coefficient, with values as low as about 0.2 and as high as
approximately 0.55 reported between experiments.
We attribute the three-dimensional character of the experiments as the major cause
of the difference in lift as well as the fact that movement of the cylinder relative
to the main flow direction cannot be completely avoided in experiments.

\section{Industrial implications}
\label{sec:industimpl}

We have demonstrated that it is possible to compute reliable lift coefficients
up to a Reynolds number of $10^4$.
Thus, it is natural to ask to what extent this could be of use in practical applications.
Here, we consider the example of designing bladeless wind generators.
We do this by considering some examples as summarized in Table \ref{tabl:somexapml}.
We make no claim that the discussion here is an in-depth assessment of the challenges
of designing bladeless turbines.
But our data allows a superficial assessment of some key parameters, such as Strouhal
period and resulting frequency of the vortex flow.
We review briefly research on bladeless turbines in section \ref{sec:prestud}.

For specificity, we pick cylinders of diameter $L$ ranging from one millimeter in diameter
up to $L=10$cm.
The larger diameter corresponds to about 4 inches and is a common dimension for plastic pipe.
Steel cylinders of diameter 1mm and length 1 meter can be obtained commercially.
For reference, one sixth of a centimeter is approximately one sixteenth of an inch.
Of course, other diameters may be of interest for a variety of reasons.
We have picked ones which can be obtained commercially.

Consider a temperature of 98 degrees F; at this temperature, the kinematic viscosity
of air is about 1/6 centimeters squared per second.
The temperature of 98 degrees F corresponds to a cool summer afternoon in West Texas, 
and this choice simplifies the resulting numbers.
Thus, the Reynolds number for a flow speed $U$ is Re $=UL/\nu\approx 6 L U$ provided $U$ is
also measured in centimeters per second.
In these units, the time unit is $L/U$ seconds.
The Strouhal period is about 10 non-dimensional time units for the Reynolds numbers
of interest here.
For specificity, we have interpolated the data for the Strouhal period from \cite{lrsBIBkl}.

If the flow rate is one meter per second (the speed of a slow walk), then $U=100$.
The flow rate of two meters per second  corresponds to a walk that is more brisk, and $U=200$.
A flow rate of ten meters per second is near the record speed for the 100 meter dash,
and corresponds to a fresh breeze on the Beaufort scale.

\begin{table}
   \centering
   \caption{Reynolds numbers (Re), physical time $t$ in seconds, Strouhal period $p_S$ 
(interpolated from \cite{lrsBIBkl}),
and frequency of oscillation for an untethered cylinder of length $L$ and wind speed $U$.
The column mph gives $U$ in miles per hour for convenience.}
\label{tabl:somexapml}
   \vspace*{5pt}
   \begin{tabular}{ccccccc}
    \toprule
$L$     &   $U$  &  mph &  Re  &   $t$   & $p_S$ & $f$ \\
\midrule
1mm     & 2 m/s  & 4.47 &  120 & 0.0005  & 11.3  & 177 Hertz \\
5/6 cm  & 1 m/s  & 2.24 &  500 & 8.33e-3 &  8.82 & 13.6 Hertz \\
1/6 cm  & 10 m/s & 22.4 & 1000 & 1.67e-4 &  8.36 & 718 Hertz \\
5/6 cm  & 2 m/s  & 4.47 & 1000 & 4.16e-3 &  8.36 & 28.7 Hertz \\
10/6 cm & 2 m/s  & 4.47 & 2000 & 8.33e-3 &  8.4  & 14.1 Hertz \\
5/6 cm  & 10 m/s & 22.4 & 5000 & 8.33e-4 &  9.0  & 133 Hertz \\
10cm    & 1 m/s  & 2.24 & 6000 &   0.1   &  9.1  &  1.1 Hertz \\
\bottomrule
   \end{tabular}
\end{table}

We highlight some information from Table \ref{tabl:somexapml}.
First, we see many examples where cylinders of practical size could be used
to generate energy in winds that commonly occur.
Second, we see that the frequency of oscillation of the cylinders would be
so fast that the movement might be seen as a blur.

We have not tried to assess how wind-induced vibrations could be used
to generate electricity.
We should recall that the force due to the lifting vibrations scales quadratically
with wind speed; the coefficient of that quadratic term is the quantity plotted
in Figure \ref{fig:liftdata}.
Thus different approaches may be needed at different wind speeds to capture the
energy generated, since the force associated with Table \ref{tabl:somexapml}
varies by 4 orders of magnitude.

Finally, note that the data in Figure \ref{fig:liftdata} and
Table \ref{tabl:somexapml} will likely change for vibrating cylinders.
The {\tt octave} file containing the exact numbers in Figure \ref{fig:liftdata}
is available upon request to the corresponding author.

\section{Studies of vortex-induced vibrations}
\label{sec:prestud}

To study vibrating cylinders, it is necessary to do more complex simulations
with moving boundaries 
\cite{vcanic2021moving,
vcanic2026existence,
chung2025stabilization,
ghattas1995variational,
kuhl2003arbitrary,
tezduyar2001finite}.
One example of such work is \cite{ref:finitevolumeVIVloRe}, which implements
 codes in OpenFoam.
They present data at Reynolds number 100.

OpenFoam may not be the ideal discretization approach for low Reynolds numbers.
In \cite{ref:KornbleuthFoamCylinder}, Strouhal numbers and frequencies
computed with OpenFoam were found to be in error by greater than a factor of two
for $Re\in[55,161]$.

The paper \cite{ref:uranssstVIV} uses turbulence models with several parameters to
simulate flow at Re=150.
Table 1 in \cite{ref:uranssstVIV} lists ten turbulence coefficients for the $k-\omega$ SST model.
They introduce a parameter $U$ that is the ratio of the incoming wind speed
to the cylinder diameter times the natural (lowest) vibration frequency of the cylinder.
They implement codes in OpenFoam.

The use of electromagnetic wind energy harvesters based on flow-induced vibrations
for powering wireless sensor networks is reviewed in \cite{ref:wirelessensorFIV}.
In \cite{ref:triboelectricVIV},  the triboelectric mechanism for electric power extraction
is detailed.
In \cite{chen2024viv}, the performance of a $2\times 2$ array of piezoelectric
harvesters (PEH) based on wind-induced vibration was measured.
The paper \cite{ref:MREbasedBWT} explores the use of magnetorheological elastomers (MREs)
to broaden the range of resonance for bladeless wind turbines.

\section{Conclusions}
\label{sec:conclusions}

We have presented lift data for flow past a cylinder for Reynolds numbers
ranging from 120 to 8000.
Lift fluctuations are typically larger than drag fluctuations as is indicated by comparing 
the horizontal and vertical scales in Figures \ref{fig:fazedataRe} and \ref{fig:fazedatacomp}.
We have compared this computational data with experimental data and obtained
good agreement in some cases, but also significant disagreement in others.
This indicates that more attention should be given to measuring lift.
Finally, we used this data to estimate the Strouhal period and frequency of the flow around cylinders of practical sizes in realistic wind speeds, showing the potential for bladeless wind turbines.
It also suggests the importance of simulating flow past an oscillating cylinder.

\section*{Acknowledgments}

Part of this research was conducted using computational resources and services at the
Center for Computation and Visualization, Brown University.
This material was initiated with support from the National Science Foundation
under Grant No. DMS-1929284 while the authors were in residence at the Institute
for Computational and Experimental Research in Mathematics in Providence, RI,
during the Numerical PDEs: Analysis, Algorithms, and Data Challenges program.

\bibliographystyle{plain}

\end{document}